\documentclass[11pt,twoside]{article}
\usepackage{aussois2003,epsf}

\def\edcomment#1{\iffalse\marginpar{\raggedright\sl#1\/}\else\relax\fi}
\reversemarginpar
\begin{document}
%
\title{Flux Ratio Anomalies: Micro- and Milli-lensing}
%
\author{Paul L. Schechter}
\affil{Massachusetts Institute of Technology, 
77 Massachusetts Avenue, Cambridge MA, 02139 USA}

\label{page:first}
\begin{abstract}
Simple models for lensing potentials that successfully reproduce the
positions of quadruple images to high accuracy fail abysmally in
reproducing the flux ratios of the multiple images, suggesting the
presence of small scale structure within the lensing galaxies.  It has
been argued that the flux ratio anomalies observed at radio
wavelengths signal the presence of CDM mini-halos.  We argue that at
least some of the anomalies observed at optical wavelengths result
from micro-lensing by stars.  This model succeeds in explaining
observed asymmetry between minima and saddle-points of the light
travel time only if a substantial fraction of the projected mass is in
a smooth, dark component.
\end{abstract}

\section{Flux Ratio Anomalies?}

There is a theorem in gravitational lensing that says, under 
certain circumstances,\footnote
{The separation between images must be small compared to the
displacement of the images from the galaxy and the gravitational
potential must be smooth on the scale of the image separation.  The
theorem is then a direct consequence of expanding the light
travel time as a power series in the neighborhood of a critical
curve, keeping terms up to third order in distance from the critical
curve.
}
a close pair of images in a quadruple system will be both
bright and of equal
brightness (e.g. Gaudi and Petters 2002).  The
archetype of such systems, PG1115+080 (Weymann et al. 1980), has a pair
of bright images separated by $0\farcs48$.

\begin{figure}
\vspace{3.0truein}
\includegraphics{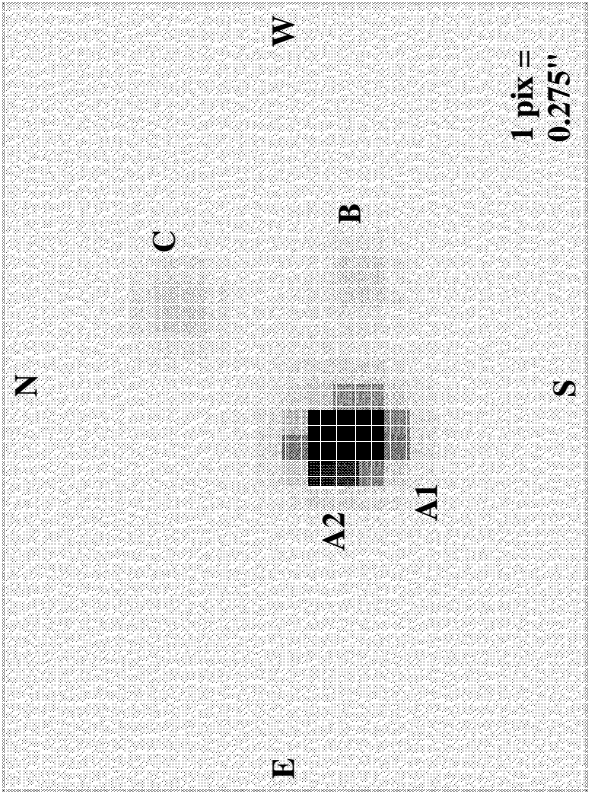}
\caption{An $I$ filter direct image of PG1115+080 taken in 0\farcs6
seeing with the 2.4-m Hiltner telescope, showing QSO components $A1,
A2, B,$ and $C$.}
\end{figure}

But as is often the case, the archetype turns out not to be
archetypical.  Zhao and Metcalf (2002) have examined a range of
``reasonable'' models for PG1115 and find that the rms magnitude
difference for the $A1$ and $A2$ (figure 1) images ought to be roughly
10\%.  The observed difference appears to have varied with time, having
ranged from 5-50\% (Vanderriest et al. 1986; Kristian et al. 1993;
Courbin et al. 1997; Iwamuro et al 2000), well outside the range
allowed by the models.  One might argue that the combination of
intrinsic variability and gravitational time delay could produce the
observed differences, but the predicted differential time delay
between $A1$ and $A2$ is of order one day and the quasar varies much
more slowly (Schechter et al. 1997).

Other quads with similar configurations have the same problem, only
worse.  MG0414+0534 has a I-filter magnitude difference $A1-A2 = 0.9$
(Schechter and Moore 1993).  By contrast the radio flux ratio (Moore
and Hewitt 1977), $A2/A1$, is very nearly unity.  Reimers et
al. (2002) call HS0810+2554 a twin to PG1115, but its bright pair of
images differ by 0.7 mag.  Most recently and most dramatically, the
system SDSS0924+0219 (Inada et al. 2003) has what appears to be a
close pair of images with a flux ratio of nearly ten.

\begin{figure}
\vspace{3.0truein}
\includegraphics{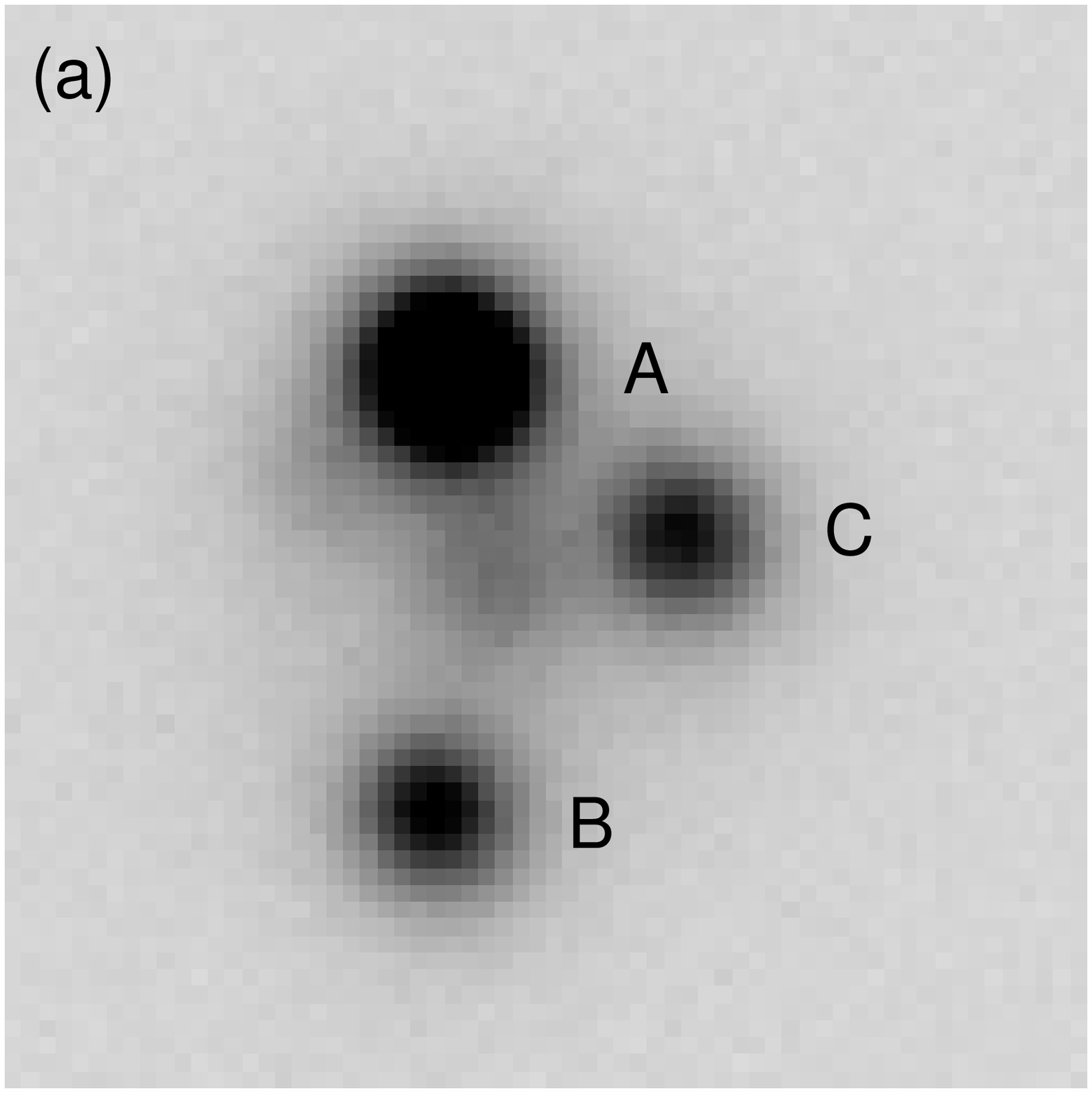}
\includegraphics{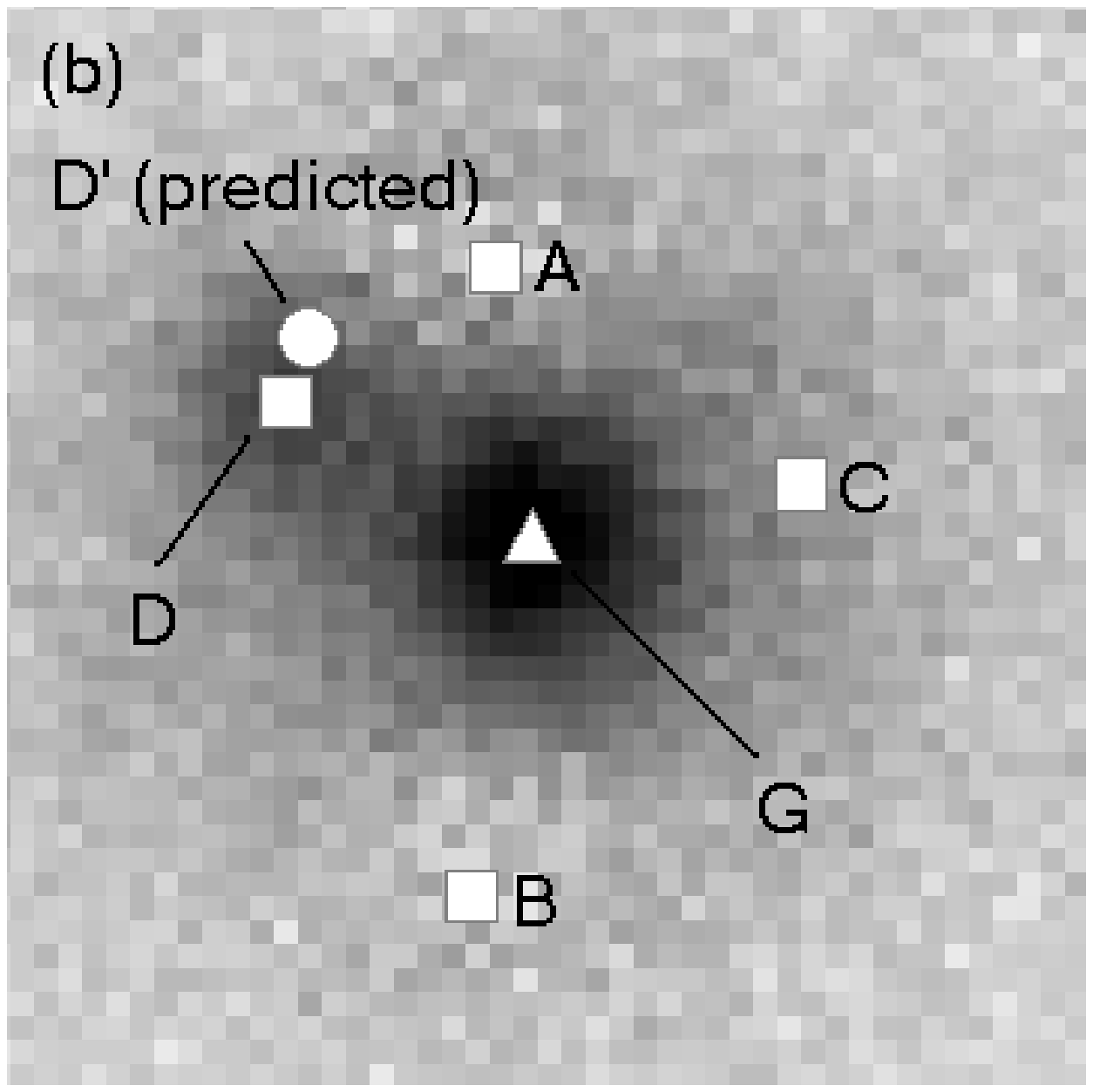}
\caption{(a) A Sloan $i'$ image SDSS0924+0219 taken in 0\farcs5 seeing
with the 6.5-m Baade telescope.  Images $A$ and $B$ are separated by
1\farcs82.  (b) A higher contrast image, with the three brightest
components subtracted.  A fourth image $D$ is visible close to the
position $D'$ predicted using the positions for images $A$, $B$, $C$
and that of the lensing galaxy $G$.  Image $A$ is a minimum while
$D'$ is a saddle-point.}
\end{figure}

Such ``flux ratio anomalies'' are not restricted to systems with close
pairs of images.  Systems with three close images and a more distant
image (e.g. B1422+231; Mao and Schneider 1998) and ``Einstein
crosses'' (e.g.  HE0435-1223; Wisotzki et al. 2002) exhibit
similar anomalies.

\section{Micro-lensing?}

The problem with the theorem (and with the models) is that they assume
a gravitational potential which is smooth on the scale of the image
separation.  With the discovery of the first lensed system Chang and
Refsdal (1979) predicted flux variations due to micro-lensing by the
stars in the intervening galaxy.  The associated graininess in the
gravitational potential will affect the fluxes if the Einstein rings
of the stars are larger than the source.

Witt, Mao and Schechter (1995) attempted to explain the anomalous
fluxes in MG0414 as the result of micro-lensing.  They took the galaxy
to be comprised entirely of stars (as opposed to an admixture of stars
and dark matter) so as to maximize the effects of micro-lensing.  The
observed $A2/A1$ ratio was at the limits of what might reasonably be
expected.  But the flux ratio in Inada's new system, SDSS0924, is more
extreme than in MG0414, well outside the anything seen in the Witt et
al. simulations.

Understandable though it might have been, Witt et al.\ were mistaken
in thinking that a galaxy comprised entirely of stars would maximize
micro-lensing fluctuations.  Deguchi and Watson (1987) and Seitz,
Schneider and Wambsganss (1994) had shown that for the case of zero
shear, increasing the optical depth to micro-lensing beyond a certain
point {\it decreased} the rms amplitude of the micro-lensing
fluctuations.  More recently, Schechter and Wambsganss (2002) have
shown that for systems like MG0414, micro-lensing fluctuations are
enhanced by keeping the surface mass density constant but substituting
smooth (and presumably) dark matter for some (but not all) of the
stellar micro-lenses.  A case as extreme as that of SDSS0924 is no
longer impossible.

\begin{figure}
\plotone{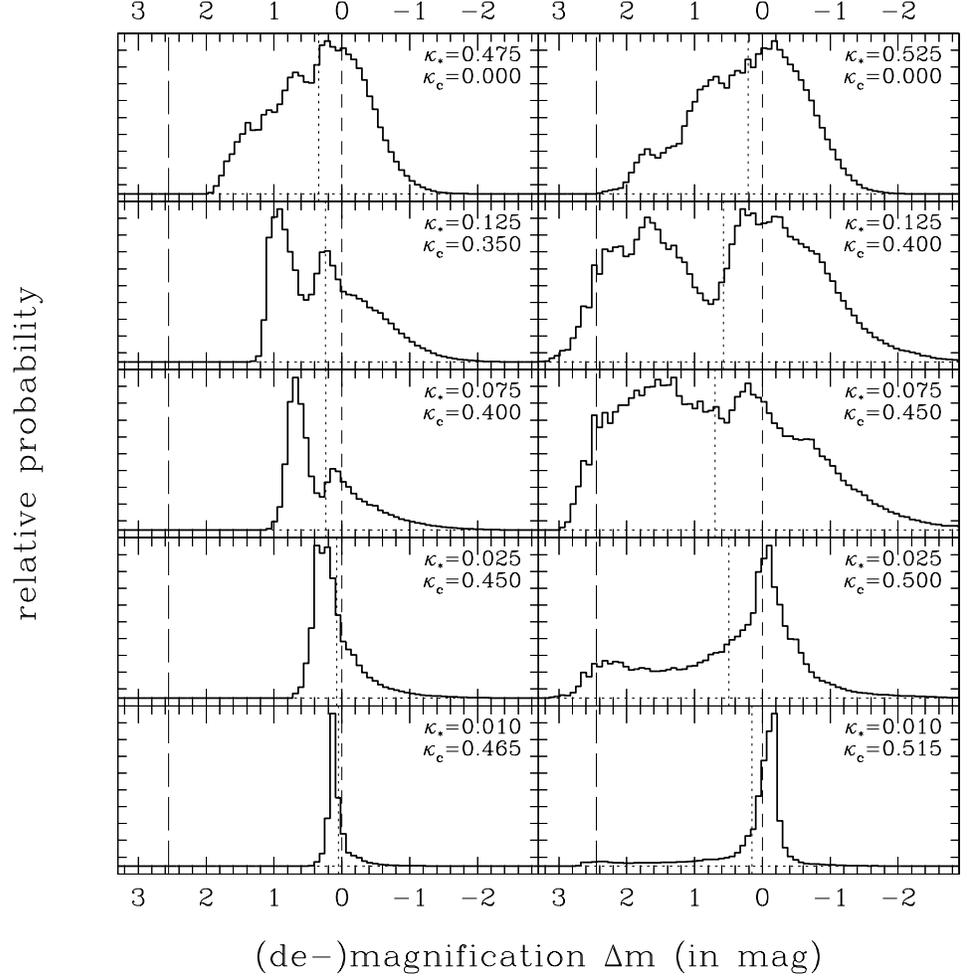}
\caption{Magnification probability distribution for a minimum (left)
with $(\kappa_{tot},\gamma) = (0.475,0.425)$ and a saddle-point
(right) with $(\kappa_{tot},\gamma) = (0.525,0.575)$, both with
magnification $\mu \approx 10$. The total convergence, $\kappa_{tot}$,
remains constant for each column.  The smoothly distributed matter
increases from top to bottom, with fractional contributions of 0\%,
75\%, 85\%, 95\% and 98\%, respectively.  The three vertical lines
indicate the following: short-dashed: $\Delta m = 0$ mag
(theoretically expected macro-magnification, $\mu \approx 10$);
dotted: $\left<\Delta m\right>$ (average magnification in magnitudes);
long-dashed: $\mu_{\rm abs} = 1.0$ (absolute magnification unity, i.e.
unlensed case).}
\end{figure}

This counterintuitive result may be explained by a two part argument.
First, at high magnification, a screen of micro-lenses produces a
large number of extra positive parity micro-images (Paczy\'nski 1986;
Wambsganss, Witt and Schneider 1993; Granot, Schechter and Wambsganss
2003).  Negative parity micro-images can (and mostly do) have
magnifications less than unity, but positive parity micro-images must
have unit magnification or greater.  As the number of positive parity
micro-images grows large, the fluctuations drop as the square root of
the number.  On the other hand, at very low optical depth,
fluctuations are rare and the rms must be small.  One might reasonably
expect the fluctuations to be largest when the number of extra positive
parity micro-images is of order unity.

Paczy\'nski (1986) has shown that for a random distribution of
micro-lenses with convergence $\kappa_*$, immersed in a smooth mass
sheet with convergence $\kappa_c$ and under the influence of an
external shear $\gamma$, there is an equivalent configuration with no
smooth component, but with effective convergence $\kappa_*^{eff} =
\kappa_*/(1-\kappa_c)$ and effective shear $\gamma^{eff} =
\gamma/(1-\kappa_c)$, subject to the condition that magnifications
computed in the effective model, $\mu^{eff}$, must be multiplied by
$(1-\kappa_c)^{-2}$.  
By adjusting the relative
contributions of $\kappa_*$ and $\kappa_c$ to the value of
$\kappa_{tot}$ determined by the smooth lens model, one changes
the number of extra positive parity micro-images.

Schechter and Wambsganss found (figure 3) that for values of
$\kappa_{tot}$ and $\gamma$ typical of PG1115-like systems, a dark
matter fraction of 50-90\% of the surface density maximizes the
micro-lensing fluctuations.  They argue that such a dark matter
fraction is consistent with the mass to light ratios observed for
elliptical galaxies.

They also found that diluting a 100\% grainy surface density with a
smooth component affected positive and negative parity macro-images
differently.  In the case of positive parity macro-images (minima of
the light travel time), substituting a smooth dark component
introduces a lower limit on the combined flux of the
micro-images,\footnote{a consequence of the requirement that there be
at least one positive parity image, and that the minimum magnification
for a positive parity image is $(1-\kappa_c)^{-2}$} eventually
narrowing the magnification distribution.  By contrast negative parity
macro-images (saddle-points of the light travel time) are quite
vulnerable to demagnification by micro-lensing.  A high magnification
saddle-point is easily split into two low-magnification micro-saddles
(Chang and Refsdal 1979; Schechter and Wambsganss 2002).  The net
result is that for pairs of bright macro-images in PG1115-like
systems, the saddle-point is almost always fainter than the than the
associated minimum.

In systems like PG1115, each image in the bright pair has a
magnification of order 10.  At the positions of the images, the shear
$\gamma$ and the total convergence and shear $\kappa_{tot}$, are
determined by the smooth lens model and are both roughly 0.5. But the
relative contributions of stars, $\kappa_*$ and dark matter,
$\kappa_c$ to the total convergence are unknown.  Were the mass all in
stars, the number of extra positive parity micro-images would be of
order 3.  If 80\% of the convergence were in a smooth component, the
effective magnification would be roughly 3 and the number of extra
positive parity images would be of order unity.

In all of the above mentioned systems (save the case of HS0810, where
the parities of the images are ambiguous) the saddle-points are
fainter than predicted in the smooth models.  The minimum/saddle-point
asymmetry had been earlier noted in the theoretical work of Metcalf
and Madau (2001), who saw differences in their cumulative
magnification distributions for their saddle-points and minima, and in
Witt et al. (1995) in their treatment of MG0414.

The foregoing considerations point to what would appear to a neat
method for measuring the relative contributions of grainy and smooth
matter to the surface mass density of lensing galaxies.  One would
assemble a ``fair'' sample of quadruple systems, model them using
image positions (but not fluxes) as constraints, and adjust the ratio
of grainy to smooth matter so as to match the observed distribution of
flux residuals.

\section{Milli-lensing!}

Astrophysics is cursed with a surfeit of explanations.  In the present
case there is second possible source of graininess.  N-body
simulations of the hierarchical clustering of cold dark matter produce
large numbers of ``mini-halos'' within the halos of typical galaxies
(Moore et al. 1999; Klypin et al. 1999).  These can be reconciled with
the much smaller numbers of dwarf satellites only if a mechanism is
invoked to prevent the baryonic matter from forming stars in these
mini-halos.  But the mini-halos (which produce deflections of order a
milliarcsecond) might still be expected to cause milli-lensing of
background QSOs.

How then, might one distinguish between milli- and micro-lensing?
Kochanek and Dalal (2003), following Koopmans and de Bruyn (2000),
argue that quasar radio source sizes are expected to be larger than
the Einstein rings of stars, ruling out micro-lensing as the source of
radio flux ratio anomalies.  The lensed system B1555+375 (Marlow et
al. 1999) is a radio analog of PG1115, whose bright components have a
flux ratio of 0.57. Kochanek and Dalal analyze a sample of radio
quads and find that the brighter of the two saddle-points is
significantly fainter than the model predictions.  

Dalal and Kochanek (2002) argue that roughly 2\% of the projected mass
density (with a factor of three uncertainty) must be in the form of
mini-halos.  Evans and Witt (2002) argued that multipoles of higher
order than quadrupole might produce similar results, but Kochanek and
Dalal argue that these would require galaxies far more misshapen than
the observed lenses.

\section{Micro- {\it and} Milli-lensing!}

While micro-lensing cannot explain the radio anomalies, milli-lensing
{\it can} explain the optical anomalies.  Is there the {\it any} need
for micro-lensing?

Two separate lines of argument suggest that micro-lensing is also
important.  First, the broad emission line regions of quasars are
thought to be larger than the Einstein rings of micro-lenses
(e.g. Moustakas and Metcalf, 2002).  An absence of broad emission
line anomalies would argue for micro-lensing.  Indeed differences
between emission-line and continuum flux ratios in several systems
(Wisotzki et al. 1993; Schechter et al. 1998; Burud et al. 2002).
Second, micro-lensing produces uncorrelated variations in multiple
images.  These have been seen, most famously in Huchra's lens,
B2237+0305 (Wo\' zniak et al. 2000) but also in HE1104-1805 (Schechter et
al. 2003) and B0957+561 (Schild 1996; Refsdal et al. 2000), and
with less coverage, in B1600+434 (Burud et al. 2001) and several other
systems.  The timescale for milli-lensing variations would be
thousands of years.  Thus it seems that both micro- and milli-lensing
are at work.

How much of the discrepancy between the observed optical fluxes and
the models is due to micro-lensing and how much is due to
milli-lensing?  Ideally one would observe lenses in both the radio and
optical and compare flux ratios.  The difference between the two would
be attributable to micro-lensing.  Unfortunately 90\% of quasars are
radio quiet.  But with the advent of integral field units (and with
judicious use of HST) one can imagine obtaining broad emission line
flux ratios for a large sample of quads.  We may therefore soon know
whether micro- or milli-lensing is the major contributor to the
observed anomalies.  When we do we will have measures of the relative
contributions of smooth and clumpy dark matter to the mass budget of
galaxies.

\acknowledgements I am indebted to my colleagues Scott Burles,
Jonathan Granot, Naohisa Inada, Joachim Wambsganss and Lutz Wisotzki,
without whose sustained efforts there would be no story to tell.

\label{page:last}
\end{document}